\pdfoutput=1
\documentclass[
    aps,pre,twocolumn, reprint,
    superscriptaddress,
    floatfix, amssymb,longbibliography
]{revtex4-2}
\usepackage[english]{babel}

\usepackage{amsmath}
\usepackage{amsfonts}
\usepackage{graphicx}
\usepackage{bm}

\usepackage[utf8]{inputenc}

\usepackage{nicefrac}

\usepackage[normalem]{ulem}

\usepackage{xcolor}

\definecolor{linkColor}{rgb}{0,0.3,0.7}
\usepackage{hyperref}
\hypersetup{colorlinks=true,
            allcolors=linkColor,
            pdfborder={0 0 0},
            pdfencoding = auto
            }

\begin{document}

\title{Spontaneous mechanical patterning in active elastic tissues with stress feedback}

\author{Henrik Weyer}
\email{henrik.weyer@kitp.ucsb.edu}
\thanks{H.W.\ and D.S.\ contributed equally to this work.}
\affiliation{Kavli Institute for Theoretical Physics, University of California Santa Barbara, Santa Barbara, California 93106, USA}

\author{Divyoj Singh}
\email{divyoj@ucsb.edu}
\thanks{H.W.\ and D.S.\ contributed equally to this work.}
\affiliation{Department of Physics, University of California Santa Barbara, Santa Barbara, California 93106, USA}

\author{M. Cristina Marchetti}
\email{mcmarche@physics.ucsb.edu }
\affiliation{Department of Physics, University of California Santa Barbara, Santa Barbara, California 93106, USA}

\date{November 14, 2026}
	
\begin{abstract}
    The active mechanics of biological tissues is often governed by embedded active contractile fibers responding to mechanical feedback.
    We introduce a minimal model of a solid-like tissue where the mechanics is controlled by feedback of local active and elastic stresses on fiber organization.
    The resulting tissue generically exhibits a bistable response to external strain and spontaneously segregates into domains of orthogonally aligned fibers.
    Our theory thus provides a minimal mechanical feedback mechanism generating orthogonal structural alignment ubiquitous in biological tissues.
\end{abstract}

\maketitle

Animal development gives rise not only to complex shapes but also to highly organized tissue architectures.
For instance, muscle fibers are often highly aligned and, in many organs, form distinct, mutually orthogonal domains that provide structural reinforcement or enable directional contractility.
Examples are the smooth-muscle layers of the gut \cite{Huycke.etal2019} or the two-layer epithelium of the fresh water polyp \emph{Hydra}~\cite{Maroudas-Sacks.etal2021,Aufschnaiter.etal2017}.
The remodeling and structural patterning in these systems are inherently mechanical and are driven by cell-scale activity such as contractile stresses generated by actomyosin fibers~\cite{Heisenberg.Bellaiche2013,Streichan.etal2018}.
Mechanical feedback guides the dynamics onto consistent trajectories, ensuring reliable development of shape and order~\cite{hannezoMechanochemicalFeedbackLoops2019,Rombouts.etal2023,petridouMultiscaleForceSensing2017,Noll.etal2017,gustafsonPatternedMechanicalFeedback2022a,Huycke.etal2019,Caldarelli.etal2024}.
Thus, the feedback between mechanical stresses in the cellular and extracellular matrix, activity, and orientational order plays a central role in organizing the dynamics and target structure of the tissue~\cite{serraMechanochemicalModelRecapitulates2023,Brauns.etal2024,perez-verdugoMechanosensitiveFeedbackOrganizes2026,Dye.etal2021}.

The interplay between alignment and elasticity has previously been studied to understand how cells align in response to external strain and interact with each other through an elastic substrate \cite{Schwarz.Safran2013,Bischofs.Schwarz2003,De.etal2007,De.Safran2008,Zemel.etal2006,Bischofs.etal2004,Bischofs.Schwarz2005,Friedrich.Safran2011,Zemel.etal2010} and the folding of prepatterned cell sheets \cite{Guillamat.etal2026}.
A similar mechanism has been successfully employed to model the endothelial tissue layer of pressurized vessels~\cite{Dessalles.etal2025} and to simulate aspects of the muscle-fiber architecture in the heart~\cite{Kroon.etal2009,Pluijmert.etal2012,Mehri.etal2024}.
Theoretical work has shown that the interplay between activity, nematic order, and elasticity leads to (quasi-)long-range order, instabilities~\cite{Maitra.Ramaswamy2019, Yin.Mahadevan2023,Roychowdhury.etal2025,Yang.etal2025}, defect motion~\cite{Brauns.etal2026}, and can dramatically alter tissue rheology in a vertex model~\cite{yu2025feedback}.

\begin{figure}
    \centering
    \includegraphics[]{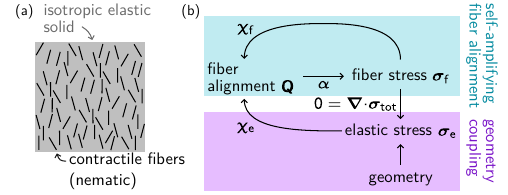}
    \caption{
    Mechanical feedback in active elastic tissues.
    (a) Schematic of the continuum model of an elastic tissue (gray) with anisotropic active contractility (embedded fibers, black) described by a nematic field $\mathbf{Q}$.
    (b) Mechanical feedback is introduced by coupling fiber alignment to the elastic and fiber stresses $\tilde{\bm\sigma}_\mathrm{e,f}$ with coefficients ${\chi_\mathrm{e,f}>0}$.
    Fiber-stress feedback induces self-amplifying fiber alignment (cyan) while the feedback from elastic stress couples the fiber orientation to tissue geometry (magenta).
    }
    \label{fig:model}
\end{figure}

Here, we demonstrate that mechanical feedback between stresses and fiber alignment in an active elastic tissue generically induces \emph{both} structural and mechanical bistability, patterning the tissue into domains of orthogonally aligned fibers.
Specifically, we introduce a minimal model of a solid-like tissue that couples active contractile fibers to cellular and extracellular matrix (ECM) elasticity [Fig.~\ref{fig:model}(a)]. We show that spontaneous nematic alignment of fibers and their structural patterning arises from a minimal mechanical-feedback law, whereby fiber alignment is driven by both active fiber and elastic stresses [Fig.~\ref{fig:model}(b)]~ \cite{Weitkunat.etal2014,Kolley-Kochel.etal2025,Ioratim-Uba.etal2023,Dye.etal2021,Mehri.etal2024}.
Thus, our theory presents a minimal picture of how anisotropic active force generation on the cellular level can coordinate via elastic coupling to mechanically pattern a tissue.

\textit{Continuum tissue model.\;---}
We approximate the background tissue (for instance, isotropic cytoskeletal structures and ECM) as a linear isotropic elastic solid [Fig.~\ref{fig:model}(a)].
This simplification captures the roughly elastic propagation of forces through the tissue, neglecting tissue flow and remodeling. We quantify deformations in terms of
the linearized strain tensor ${\bm\varepsilon=\tfrac{1}{2}[ \bm\nabla\mathbf{u}+(\bm\nabla\mathbf{u})^\mathsf{T}}]$, with $\mathbf{u}(\mathbf{x},t)$ the displacement field. The elastic background material is then described by the stress--strain relation $\bm\sigma_\mathrm{e} = 2\mu \Tilde{\bm\varepsilon} + B \mathrm{Tr}[\bm\varepsilon] \mathbf{1}$, with $\mu,B$ the shear and bulk moduli and $\mathbf{1}$ the identity matrix.
Deviatoric (trace-less) quantities are denoted by a tilde.

Active fibers embedded in this isotropic background are described by a nematic symmetric and traceless tensor field $\mathbf{Q}(\mathbf{x},t)$ and exert a contractile fiber stress ${\bm\sigma_\mathrm{f}=\alpha\mathbf{Q}}$, where ${\alpha>0}$ is the active contractility~\cite{kruse2005generic} [Fig.~\ref{fig:model}(a)].
Following the Landau--de Gennes expansion \cite{Gennes.Prost2013}, we describe the dynamics of fiber remodeling by
\begin{subequations}
\begin{align}
    \partial_t \mathbf{Q} &= - (a + b\, \mathrm{Tr}[\mathbf{Q}^2])\mathbf{Q} + K \nabla^2 \mathbf{Q} + \chi_\mathrm{f} \tilde{\bm\sigma}_\mathrm{f} + \chi_\mathrm{e} \tilde{\bm\sigma}_\mathrm{e}\label{eq:fiber-dyn}\\
    &= (\chi_\mathrm{f}\alpha - a) \mathbf{Q} - b \mathrm{Tr}[\mathbf{Q}^2]\mathbf{Q} + K \nabla^2 \mathbf{Q} + \chi_\mathrm{e} \tilde{\bm\sigma}_\mathrm{e}\,.\label{eq:fiber-feedback}
\end{align}
\end{subequations}
The first term in Eq.~\eqref{eq:fiber-dyn} models the decay of unloaded fibers (see, e.g., Ref.~\cite{Tojkander.etal2015}), with ${a,b>0}$.
The nonlinear term ensures the saturation of the magnitude of $\mathbf{Q}$, focusing on two-dimensional systems here.
The Frank-elasticity term in the approximation of a single elastic constant $K$ requires neighboring cells to align fibers in the same direction, penalizing gradients in fiber orientation.
This term is phenomenological. An interesting open question is to derive it from a cell-scale tissue model.
The last two terms in Eq.~\eqref{eq:fiber-dyn} describe the coupling to fiber and elastic deviatoric stresses and constitute the mechanical feedback.
This choice of feedback is motivated by the observed maturation of muscle fibers under tension \cite{Weitkunat.etal2014,Kolley-Kochel.etal2025}. It has been used to analyze cell-shape patterning based on cell rearrangements \cite{Dye.etal2021}, and it arises in a model of tissue-scale feedback caused by actomyosin catch-bonds  \cite{Ioratim-Uba.etal2023}.
Importantly, the feedback acts on the cellular scale, where the fiber and elastic stresses are carried by different molecular structures and can therefore induce distinct cellular responses.
Moreover, even coarse-graining force dipoles aligning with respect to local strain (elastic stress) results in alignment with respect to both elastic \emph{and} fiber stresses at the continuum level \cite{Friedrich.Safran2011}.
We therefore introduce two independent coupling parameters ${\chi_\mathrm{e,f}>0}$.
For simulations, we choose the symmetric coupling ${\chi_\mathrm{e}=\chi_\mathrm{f}\equiv \chi}$ which corresponds to a coupling to total stress ${\bm\sigma_\mathrm{tot}=\bm\sigma_\mathrm{e}+\bm\sigma_\mathrm{f}}$.
Finally, we assume the tissue to be in mechanical equilibrium, ${\bm\nabla\cdot\bm{\sigma}_\mathrm{tot}=0}$, because fiber remodeling occurs on timescales much longer than mechanical relaxation (see, e.g., Refs.~\cite{Noll.etal2020,Brauns.etal2024}).

\begin{figure}
    \centering
    \includegraphics[]{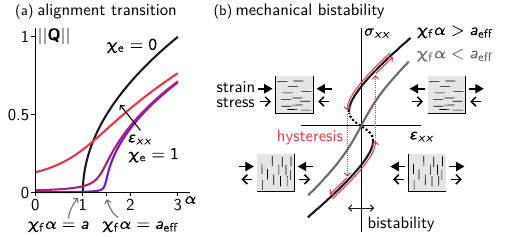}
    \caption{
    Fiber transition and mechanical bistability.
    (a) Without elastic coupling (${\chi_\mathrm{e}=0}$, black), the fiber-stress feedback introduces spontaneous fiber alignment ${||\mathbf{Q}||>0}$ if ${\chi_\mathrm{f}\alpha>a}$. 
    Elastic-stress feedback increases the alignment threshold to ${\chi_\mathrm{f}\alpha=a_\mathrm{eff}}$. 
    The alignment then depends on the mechanical boundary condition, shown here by increasing the uni-axially imposed strain ${\varepsilon_{xx}=0.001,0.01,0.1}$ (purple, magenta, red).
    The simulation parameters are ${a=b=K=\mu=\chi_\mathrm{f}=1}$ and ${B=2}$.
    (b) The self-alignment transition ${\chi_\mathrm{f}\alpha>a_\mathrm{eff}}$ introduces a bistable, hysteretic stress response $\sigma_{xx}$ (black) to uniaxial tissue strain $\varepsilon_{xx}$.
    Below the alignment transition the response is nonlinear but not bistable (gray).
    The insets show the fiber orientation, tissue stresses, and strains in the four quadrants.
    }
    \label{fig:bistability}
\end{figure}

\textit{Alignment transition.\;---}In Eq.~\eqref{eq:fiber-feedback}, we inserted the fiber stress $\alpha\mathbf{Q}$.
The coupling to fiber stress ($\chi_\mathrm{f}\not=0, \chi_\mathrm{e}=0$) induces the self-amplifying growth of contractile fibers for ${\chi_\mathrm{f}\alpha>a}$ [Fig.~\ref{fig:model}(a), black line].
Under this condition, spontaneously aligned fibers pull strongly on their neighbors.
Feedback-induced alignment then overcomes the spontaneous decay rate $a$, resulting in growth of order $||\mathbf{Q}||$. 
The additional elastic-stress feedback ($\chi_\mathrm{f}, \chi_\mathrm{e}\not=0$) provides a sensing of the system geometry through the boundary conditions that determine the local elastic stresses as the tissue deforms.
Simultaneously, the elastic stress induced by fiber contractility opposes alignment and increases the alignment threshold. 
This is shown in Fig.~\ref{fig:bistability}(a) by increasing the uni-axially imposed strain $\varepsilon_{xx}$, resulting in the alignment threshold ${\chi_\mathrm{f}\alpha=a_\mathrm{eff}}$, with $a_\mathrm{eff}=a+\chi_\mathrm{e}\alpha\mu/(B+\mu)$ derived below. 

\textit{Mechanical bistability.\;---}
The self-amplifying growth of the contractile fibers induces a bistable stress--strain relation of the composite tissue [Fig.~\ref{fig:bistability}(b)].
To see this, imagine stretching a homogeneous tissue patch adiabatically along $x$. Mechanical feedback aligns fibers along the stretching direction.
The magnitude of alignment $Q_{xx}$ can be obtained from Eq.~\eqref{eq:fiber-feedback} after determining $\tilde{\bm\sigma}_\mathrm{e}$ by solving ${\bm\nabla\cdot\bm{\sigma}_\mathrm{tot}=0}$ using the imposed boundary conditions. 
The stress-free boundary conditions in the $y$ direction imply ${\sigma_{yy}=0}$.
Using this to eliminate the strain $\epsilon_{yy}$, and assuming the alignment of fibers in $x$ direction, one obtains ${\tilde{\sigma}_{xx}^\mathrm{e}=\mu(\epsilon_{xx}-\epsilon_{yy})=\frac{2\mu B}{B+\mu}\epsilon_{xx}-\frac{\alpha\mu}{B+\mu}Q_{xx}}$. 
Inserting ${\tilde{\sigma}_{xx}^\mathrm{e}}$ in the fiber equation Eq.~\eqref{eq:fiber-feedback}, the steady state homogeneous value of fiber alignment is given by the solution of ${(\chi_\mathrm{f}\alpha -a_\mathrm{eff})Q_{xx}-2bQ_{xx}^3+\frac{2\chi_e B\mu}{B+\mu}\epsilon_{xx}=0}$. The self-alignment threshold in the presence of elastic feedback is then ${\chi_\mathrm{f}\alpha=a_\mathrm{eff}}$ [cf.\ Fig.~\ref{fig:bistability}(a)]~\cite{Article2026}. 
For ${\chi_f\alpha>a_\mathrm{eff}}$, both a negative and a positive stable solution for $Q_{xx}$ exist, separated by an unstable central fixed point.
Due to the exerted fiber stress this structural bistability implies that the stress--strain relation becomes bistable as well
[Fig.~\ref{fig:bistability}(b) and Ref.~\cite{Article2026}].
Large strain aligns the fibers but the self-amplifying feedback of fiber growth implies that these fibers remain aligned when the tissue is slowly returned into the unstrained state ${\varepsilon_{xx}=0}$.
A finite amount of compression is required to realign the fibers in the perpendicular direction, resulting in hysteresis.
While our analysis assumed fibers to be either aligned or orthogonal to the strain direction $x$, a stability analysis of transverse perturbations shows that the aligned and perpendicular fiber states are indeed stable against fiber rotation for small applied strain $\varepsilon_{xx}$ (see Ref.~\cite{Article2026}).
Moreover, if one has ${\chi_\mathrm{f}\approx\chi_\mathrm{e}}$ and the background material is nearly incompressible, both branches of the stress--strain relation are stable against homogeneous fiber rotation over their whole range of existence.
Taken together, the tissue shows mechanical bistability because of the positive feedback loop between fiber alignment and fiber stress.

\begin{figure}
    \centering
    \includegraphics[]{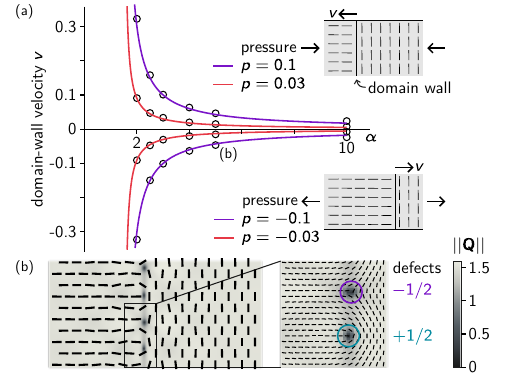}
    \caption{
    Domain-wall formation in quasi one-dimensional tissue.
    (a) A rectangular tissue patch subject to a horizontal pressure $p$ with periodic boundary conditions in the vertical direction (insets) segregates into horizontal and vertically aligned domains. 
    Equation~\eqref{eq:quasi-1D-dw-vel} (lines) predicts the domain-wall velocity observed in the two-dimensional simulations (circles) for different pressures and fiber contractilities $\alpha$.
    (b) The magnitude of alignment $||\mathbf{Q}||$ (grayscale map) and the director fields show that the domain wall decomposes into a chain of $\pm 1/2$ orientational defects in the simulations for ${\alpha>2}$; shown for ${\alpha=5}$ and ${p=-0.1}$.
    The simulation domain has dimensions ${(100,50)}$ and the model parameters are chosen as in Fig.~\ref{fig:bistability}(a).
    }
    \label{fig:quasi-1D-domainWall}
\end{figure}

\textit{Domain-wall formation.\;---}
The bistable mechanical response suggests that the fibers in extended active elastic tissues might spontaneously segregate into coexisting domains of orthogonal fiber orientations, separated by domain walls.
We study this behavior first in an effectively one-dimensional system, a tissue patch with periodic boundary conditions in the $y$ direction under applied stress, or pressure, $p$ in the $x$ direction.
Motivated by the above analysis, we again assume that the fiber direction is either parallel or orthogonal to the axis $x$ of applied load, implying ${Q_{xy}=0}$.
Then, solving the mechanical equilibrium and inserting the resulting elastic stress into the fiber dynamics, Eq.~\eqref{eq:fiber-feedback}, one obtains
\begin{equation}\label{eq:quasi-1D-AC}
    \partial_t Q_{xx} = K \partial_x^2 Q_{xx} + \tilde{a}\, Q_{xx} - 2b\, Q_{xx}^3 - \chi_\mathrm{e}\frac{\mu}{B+\mu}p
\,,
\end{equation}
with ${\tilde{a}=\chi_\mathrm{f}\alpha-a_\mathrm{eff}}$.
This equation is the well-known bistable-media or Allen--Cahn (model A) equation with a biasing term caused by the applied pressure $\sim p$ \cite{Schlogl1972,Hohenberg.Halperin1977,Allen.Cahn1979}.

Bistable media form nearly uniform domains relaxed onto the different fixed points of the homogeneous system, separated by domain walls of width ${\ell_\mathrm{dw}=\sqrt{K/\tilde{a}}}$ (Fig.~\ref{fig:quasi-1D-domainWall}).
These domain walls attain a finite velocity due to the biasing pressure term which makes one fixed point preferred over the other (lowering its effective free energy) \cite{Mikhailov1990}.
For a small magnitude of the biasing term, the domain wall moves at a velocity (see Ref.~\cite{Article2026})
\begin{equation}\label{eq:quasi-1D-dw-vel}
    v\approx- \frac{3\sqrt{K b}}{\tilde{a}}\frac{\mu}{B+\mu}\chi_\mathrm{e}p\,.
\end{equation}
For negative pressure ${p<0}$, that is, extensile stress applied to the tissue patch, alignment of fibers along the stress direction is preferred and the corresponding domain slowly takes over the whole patch [Fig.~\ref{fig:quasi-1D-domainWall}(a)].
In contrast, positive pressure ${p>0}$ induces domain-wall movement that aligns all fibers in the perpendicular direction.
Our assumption of fiber alignment parallel or orthogonal to the direction of externally applied stress breaks down in the interface region at large fiber contractility $\alpha$: The domain wall breaks up into a chain of alternating ${\pm 1/2}$ defects, the classical ``low-energy'' defects expected in a nematic system [Fig.~\ref{fig:quasi-1D-domainWall}(b)].
It is an intriguing open question what drives the transition between ${\pm 1/2}$ orientational defects and domain walls as the emergent collective degrees of freedom of the system, and what determines the spacing of the defects.
We expect that the transition stems from competition between elasticity driving the directors to orient orthogonally but increasing nematic self-alignment disfavoring line compared to point defects.
Nonetheless, Figure~\ref{fig:quasi-1D-domainWall}(a) shows that the domain-wall velocity is well described by Eq.~\eqref{eq:quasi-1D-dw-vel} in either case.

\textit{Discussion.\;---}
We showed that the positive feedback between mechanical stress and the alignment of contractile fibers provides a generic mechanism for the formation of orthogonally aligned fiber domains in biological tissue. The coupling to \emph{both} elastic and active stresses is key to engender structural and mechanical bistability - a key distinction from earlier works exploring alignment induced by elastic stresses or strains~\cite{yu2025feedback}. 
Moreover, our prediction can be directly
tested in organoids \cite{Merk.etalinpreparation} and explains observations in \textit{Hydra} \cite{Weyer.Braunsinpreparation}.

An exciting future research direction is incorporating plastic deformations and viscoelasticity of the background material, allowing for tissue remodeling through dynamical patterning of anisotropic contractility \cite{Ioratim-Uba.etal2023,Sknepnek.etal2023,Dye.etal2021}.
Moreover, an intriguing theoretical challenge is understanding how elasticity induces a transition between the domain-wall (line) defects typical of multistable systems with discrete symmetry and a chain of  $\pm1/2$ orientational defects characteristic of systems with continuous nematic symmetry.
Finally, we will next consider three-dimensional tissues where the isotropic--nematic transition becomes first order, and more than two orthogonal fiber directions could coexist.

\vspace*{1em}
We thank Fridtjof Brauns, Markus Merk and Sebastian Streichan for insightful discussions.
This work was supported by NSF award DMR-2528734.
H.W. was supported by grant GBMF7392 from the Gordon and Betty Moore Foundation and NSF PHY-2309135 to the Kavli Institute for Theoretical Physics (KITP).
This research has benefited from the Mitchel Postdoctoral Scholar Career Development Fund.

%

\end{document}


\maketitle

\section*{Numerical simulations}

\subsection*{Implementation}

The coupled dynamics of the nematic order parameter $\bm{Q}$ and the displacement field $\bm{u}$ is solved with the time-dependent finite-element solver of COMSOL
Multiphysics (version 6.2), using linear or quadratic Lagrange elements for all fields in the \emph{General Form PDE} interfaces.

\subsection*{Geometry and mesh}

The annular domain was created using concentric circles of radii
$R_+$ and $R_-$. All meshes were built using 
COMSOL's Delaunay algorithm.
The simulation of the quasi-one-dimensional tissue patch in Fig.~3 uses a linear mesh size of $0.2$.

\subsection*{Boundary and initial conditions}

The simulations of the quasi-one-dimensional tissue patch employ prescribed-pressure (traction) boundary conditions in $x$ direction, while periodic boundary conditions are used in $y$ direction.
To consistently orient the coexisting domains, the initial condition is chosen as weak noise superimposed onto the gradient ${Q_{xx}=1-2x/L}$, where the tissue covers $x\in [0,L]$.

\subsection*{Time integration}

Time stepping uses the implicit BDF solver with free time steps and orders
between 1 and 5, at relative and absolute tolerances of $10^{-3}$ or smaller (as specified in the supplied setup files).  The two
interfaces are solved together, by a fully coupled Newton iteration with the
MUMPS or PARDISO direct linear solvers. The steady states shown are the solution at ${t=1000}$.

\subsection*{Analysis of the domain wall position}

The solutions are exported on a uniform grid and the domain-wall position
$r_\mathrm{dw}$ is the interpolated zero crossing of the azimuthal mean of 
$Q_{xx}$.

\subsection*{Code availability}

The COMSOL setup files and the Mathematica and Python analysis scripts will be made available at \url{https://github.com/Divyoj-Singh/Stress_feedback_active_solid}.